# Observation of room-temperature ferroelectricity in tetragonal strontium titanate thin films on SrTiO$_3$ (001) substrates

Y. S. Kim, D. J. Kim, T. H. Kim, and T. W. Noh[a]

*ReCOE & FPRD, Department of Physics and Astronomy, Seoul National University, Seoul 151-747, Korea*

J. S. Choi and B. H. Park

*Department of Physics, Konkuk University, Seoul 143-701, Korea*

J.-G. Yoon

*Department of Physics, University of Suwon, Kyunggi-do 445-743, Korea*

We investigated the ferroelectric properties of strontium titanate (STO) thin films deposited on SrTiO$_3$ (001) substrate with SrRuO$_3$ electrodes. The STO layer was grown coherently on the SrTiO$_3$ substrate without in-plane lattice relaxation, but its out-of-plane lattice constant increased with a decrease in the oxygen pressure during deposition. Using piezoresponse force microscopy and *P-V* measurements, we showed that our tetragonal STO films possess room-temperature ferroelectricity. We discuss the possible origins of the observed ferroelectricity.

---

[a] E-mail: twnoh@phya.snu.ac.kr

SrTiO$_3$ is a well-known incipient ferroelectric (FE) material. At low temperature, it has a very large dielectric constant but remains in a paraelectric state.[1] Its FE stability seems to be hampered by competing interactions, such as quantum fluctuation and antiferrodistortion.[2,3] Such a delicate balance can easily be upset by external parameters. For example, other studies have reported ferroelectricity in SrTiO$_3$ due to doping with other cations,[4] substituting oxygen isotopes,[5] applying electric fields,[6] or using strain engineering techniques.[7-10]

In this letter, we report the observation of room-temperature ferroelectricity in tetragonal strontium titanate (STO) thin films. We grew STO thin films coherently on SrTiO$_3$ (001) substrates between SrRuO$_3$ (SRO) top and bottom electrodes. We discovered that our STO films, grown at a lower oxygen partial pressure ($P_{O2}$), have larger out-of-lattice constants and show ferroelectric behavior at room temperature. We discuss the roles of strain and defects on the observed ferroelectricity in the STO films.

To fabricate SRO/STO/SRO heterostructures on SrTiO$_3$ (001) substrate, we used pulsed laser deposition (PLD) with *in-situ* reflection high-energy electron diffraction (RHEED).[11] We used a KrF excimer laser with repetition rate of 1 Hz and fluence of ~ 2 J/cm$^2$. We deposited the bottom SRO, STO, and top SRO layers sequentially at 600°C without breaking vacuum. Before deposition, we prepared TiO$_2$-terminated SrTiO$_3$ substrates.[10] Then, we deposited an ~ 15-nm-thick SRO bottom electrode layer using a polycrystalline SrRuO$_3$ target at $P_{O2}$ of 2×10$^{-2}$ Torr. Next we grew the STO layer using a SrTiO$_3$ single crystal target at various $P_{O2}$, ranging from 2×10$^{-4}$ to 2×10$^{-2}$ Torr. We monitored the layer-by-layer growth mode carefully and obtained an ~ 100-nm-thick STO layer by counting RHEED intensity oscillations. After depositing an ~ 15-nm-thick SRO top electrode, we lowered the temperature to room temperature while keeping $P_{O2}$ constant.

We found that all of the layers, especially the STO layer, were grown coherently on SrTiO$_3$ (001) substrate. To investigate the crystal structure and lattice constants of the STO thin films, we measured x-ray reciprocal space maps (X-RSM) around the asymmetric ($\bar{1}$03) Bragg reflections. For the X-RSM measurements, we used a Bruker AXS D8 advanced x-ray diffractometer with a Vactec-1 detector. We plotted X-RSM in reciprocal lattice unit (r.l.u.), which is calculated from measured angular scans. (1 r.l.u.=2π/3.905Å$^{-1}$) Figures 1 (a) and 1(b) show the X-RSM of the SRO/STO/SRO heterostructures grown with $P_{O2}$ = 10 and 0.2 mTorr, respectively. These maps show clear (103) Bragg peaks of the SrTiO$_3$ substrate, SRO, and STO layers in the *H0L* scattering plane. All of these peaks are on the same *H*-value line, indicating that the SRO and STO layers could be grown coherently on the SrTiO$_3$ substrates without lattice relaxation.

The crystal structure of the STO layer has a systematic dependence on $P_{O2}$. Note that the difference in the *L* values of the STO layer peaks in Figs. 1(a) and 1(b) from that of the SrTiO$_3$ substrate indicates that our STO layers have tetragonal symmetry, not cubic symmetry. The X-RSM also shows that the *L* value becomes smaller for the films grown at a smaller $P_{O2}$. The solid red squares and triangles in Fig. 1 (c) show the measured values of the in-plane (a-axis) and the out-of-plane (c-axis) lattice constants, respectively. The a-axis lattice constant remains the same, independent of $P_{O2}$ due to the clamping effect of the substrate, while the c-axis lattice constant increases systematically as $P_{O2}$ decreases. The increase in the c-axis lattice constant is related to the expansion of the unit cell volume, which could be attributed to formation of defects inside the STO layer.

We define the tetragonality as ($c/a$) - 1, where $a$ and $c$ are the a- and c-axis lattice constants, respectively. In displacive ferroelectrics, such as BaTiO$_3$, ferroelectricity comes from the displacement of transition metal ions; therefore, large tetragonality is a prerequisite for ferroelectricity.[10,12] As shown in Fig. 1(c), the tetragonality of our films (the solid blue circles) increases as $P_{O2}$ decreases. The tetragonality becomes even larger than that of FE BaTiO$_3$, shown by the dashed line, especially for our STO films deposited at $P_{O2}$ of less than 10$^{-3}$ Torr.

To check whether our STO films could be FE, we performed piezoresponse force microscopy (PFM) tests. These included room-temperature PFM tests using a Park Systems XE-100 atomic force microscope (AFM) and temperature-dependent piezoresponse studies using another AFM, a Seiko Instruments SPA-300 HV. Figure 2 (a) shows a schematic diagram of our PFM study setup. First, we scanned a region 3×3 μm square with a DC bias of +10.5 V. Then, we scanned a center 1×1 μm square region with a reversed DC bias of -10.5 V. Finally, we obtained PFM images by applying an AC bias to the tip. Figures 2 (b) and 2(c) show the room-temperature phase and amplitude images of the piezoelectric domains on a STO layer that was grown at $P_{O2}$ of 0.2 mTorr. These figures indicate that read/write and reversible piezoelectric domains can be formed. Figure 2(d) shows the temperature dependence of the maximum piezoelectric coefficient, 2$d_{33max}$. This figure indicates that 2$d_{33max}$ has nearly temperature-independent values in the range of 17 to 25 pm/V.

In the configuration of Fig. 2(a), surface charges trapped on the free dielectric surface might result in spurious signals.[13] To remove any systematic errors, we performed another PFM test using our SRO/STO/SRO heterostructures, which are shown schematically in Fig. 3(a). In this configuration, the STO layer was covered with a conducting SRO top electrode, so we could avoid the surface charge trapping effects and apply a uniform electric field to the STO layer. In addition, we applied the DC and AC biases using a separate electrical connection and measured only the piezoresponse using the PFM tip. We found that this experimental setup reduced the noise significantly and provided fairly good $d_{33}$-$V$ hysteresis loops. Figure 3(b) shows a room temperature $d_{33}$-$V$ hysteresis loop for the STO layer, deposited at $P_{O2}$ of 0.2 mTorr. The observed $2d_{33max}$ value is ~ 27 pm/V. The observations of piezoelectric domains and hysteresis loops suggested that the occurrence of ferroelectricity at room temperature in our tetragonal STO thin films is genuine.

Further experimental evidence for ferroelectricity came from electrical measurements. To measure the $P$-$V$ hysteresis loops, we used a Sawyer-Tower circuit, which consisted of a Yokogawa FG300 function generator and a DL7100 digital oscilloscope, and applied triangular waves of 200 kHz to the SRO/STO/SRO capacitors. Figures 4(a) and 4(b) show the $P$-$V$ loops, measured on 50×50 μm square capacitors, for the films grown at $P_{O2}$ of 1.0 and 0.2 mTorr, respectively. These $P$-$V$ loops show typical FE responses. Figure 4(c) illustrates the temperature-dependent remnant polarization ($P_r^*$) values of our STO thin films. As the temperature decreases, the value of $2P_r^*$ increases. At 10 K, the $2P_r^*$ value of the STO film grown at $P_{O2}$ of 1.0 mTorr becomes ~ 6 μC/cm$^2$. Note that this $2P_r^*$ value is larger than the reported $2P_r^*$ values of Ca$_x$Sr$_{1-x}$TiO$_3$ (~ 0.55 μC/cm$^2$ at 4.3 K)[4] and isotope-exchanged SrTi$^{18}$O$_3$ (~ 0.6 μC/cm$^2$ at 18 K)[5]. The piezoelectric coefficient can be expressed as $d_{33} = 2 Q\varepsilon_{33}P_S$, where $Q$ is the electrostriction coefficient and $\varepsilon_{33}$ is the dielectric constant along the c-axis. Note that the temperature dependence of $2P_r^*$ in Fig. 4(c) is similar to that of $2d_{33max}$ in Fig. 2(d).

At present, we are unable to determine the origins of the room-temperature ferroelectricity in our STO films. The strain coming from the coherent growth, shown in Fig. 1, is one possible candidate. To check this possibility, we estimated the misfit strains of STO films grown at $P_{O2}$ of 1.0 and 0.2 mTorr to be ~ -3.7 and -4.3×10$^{-3}$, respectively. According to a thermodynamic calculation for stoichiometric SrTiO$_3$,[7] the misfit strains of our STO films are not large enough to raise their Curie temperatures to room temperature. Moreover, we fabricated an ~ 30-nm-thick STO film in addition to the 100-nm-thick films used in our studies. X-ray studies showed that the thinner film has nearly the same a- and c-axes lattice constants. If the room-temperature ferroelectricity comes from the strain effects only, the thinner film should have the same $2P_r^*$ values. However, as shown in Fig. 4(d), its $2P_r^*$ values are smaller than those of the 100-nm-thick films by a factor of two to three. Therefore, the strain effect alone could not be the main origin of the ferroelectricity.

Other candidates for the room temperature ferroelectricity are defects inside the STO films. In our STO films, we expect the formation of defects during deposition under lower $P_{O2}$. It has been reported that oxygen vacancies could increase the unit cell volume, consistent with our observations.[14] However, our films showed unit cell volume expansion up to 1.3%, much greater than that of SrTiO$_{3-\delta}$, which is typically less than 0.3%.[14] In addition, other studies have reported that oxygen vacancies can induce electron doping, which makes SrTiO$_{3-\delta}$ metallic.[14,15] Our STO film grown at $P_{O2}$ of 0.2 mTorr had a resistivity of ~ 10$^6$ Ohm·cm, which is much greater than that of metallic SrTiO$_{3-\delta}$. Another kind of defect comprises SrO vacancies. This possibility is supported theoretically by formation energy calculations; the energy of a SrO vacancy in SrTiO$_3$ is only 1.53 eV, which is smaller than those of other defects.[16]

Recent PLD studies of STO thin films by Ohnishi *et al.* support the formation of SrO vacancies.[17] We also have optical spectroscopy data and first-principle calculation results that are consistent with this view.[18] If the Sr vacancies can be formed with similar probability to oxygen vacancies, the doping effect will be minimal due to the neutral charge of the SrO vacancies leaving the STO films as insulators. Under the strain developed in our SRO/STO/SRO geometry, the SrO vacancies might be aligned and could provide defect dipoles in a given direction.[19] Further investigations on the formation of SrO vacancies and their roles in ferroelectricity is warranted.

In summary, we grew epitaxial strontium titanate (STO) thin films on $SrTiO_3$ (001) substrates with $SrRuO_3$ top and bottom electrodes under low oxygen pressures. The resulting tetragonal STO thin films showed ferroelectric responses even at room temperature. This interesting experimental finding is very easy to implement when making ferroelectric STO films and could have significant practical implications.

The authors thank Y. J. Chang, J. Y. Jo, S. S. A. Seo, and S. H. Chang for their valuable discussions. This work was supported financially by the Creative Research Initiatives (Functionally Integrated Oxide Heterostructure) of the Korea Science and Engineering Foundation (KOSEF).

**Figure captions**

Fig. 1. (Color online) X-ray reciprocal space mapping around the asymmetric ($\bar{1}03$) Bragg reflection of SrRuO$_3$/STO/SrRuO$_3$ heterostructure grown with oxygen partial pressure ($P_{O2}$) = 10 m Torr (a) and $P_{O2}$ = 0.2 mTorr (b). (c) The $P_{O2}$ dependences of in-plane and out-of-plane lattice constants of STO thin films are shown as solid red squares and solid red triangles, respectively. The tetragonality of STO thin films as a function of the $P_{O2}$ is shown as solid blue circles.

Fig. 2. (Color online) (a) Schematic diagram of the PFM setup. Phase (b) and amplitude (c) of the piezoelectric domain after application of +10.5 V and −10.5 V bias to the 3×3 μm and 1×1 μm areas, respectively. The PFM image was obtained at room temperature for the STO film grown at $P_{O2}$ = 0.2 mTorr. (d) Temperature dependence of the maximum piezoelectric coefficient, $2d_{33max}$ of the STO film. [$2d_{33max} = d_{33(+)} + d_{33(-)}$]

Fig. 3. (Color online) (a) Schematic diagram of another PFM setup. In this setup, uniform DC and AC bias can be applied on the SRO top electrode with a separate electrical circuit. The PFM tip only reads mechanical piezoelectric response. (b) Room temperature $d_{33}$-$V$ hysteresis loop for the STO films grown at $P_{O2}$ = 0.2 mTorr.

Fig. 4. (Color online) Temperature dependence of $P$-$V$ hysteresis loops in STO films grown at $P_{O2}$ = 1

mTorr (a) and $P_{O2}$ = 0.2 mTorr (b). (c) Temperature dependence of the remnant polarization $2P_r^*$ for STO films grown at $P_{O2}$ = 0.2 mTorr and 1 mTorr. The films are 100 nm thick. (d) Temperature dependence of $2P_r^*$ for 100- and 30-nm-thick STO films, grown at $P_{O2}$ = 0.2 mTorr. $[2P_r^* = P_{r(+)} + P_{r(-)}]$

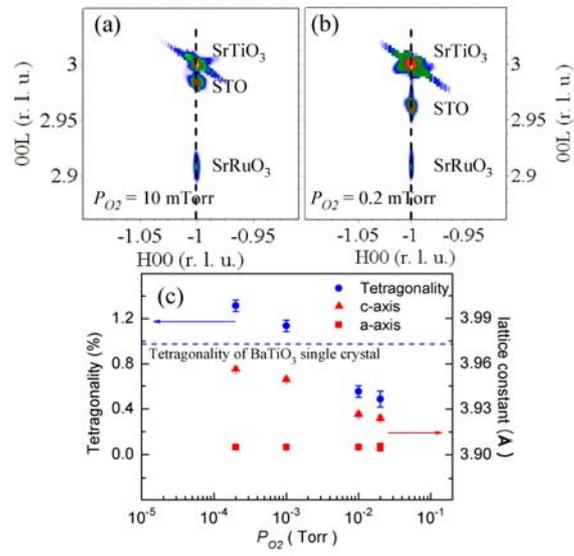

Fig. 1

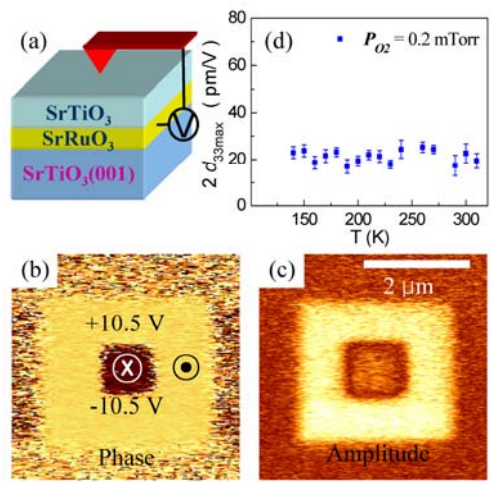

Fig. 2

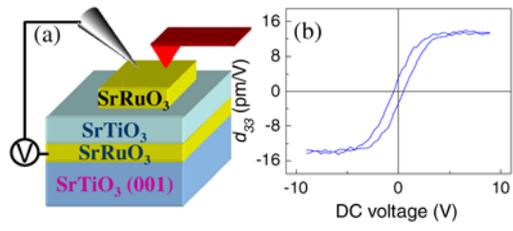

Fig. 3

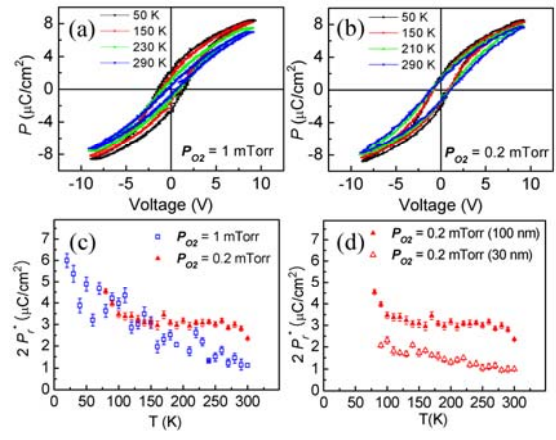

Fig. 4